\documentclass[twoside,11pt]{article}

\usepackage{jmlr2e}
\usepackage{amsmath}
\usepackage{booktabs}

\def\bX{\mathbf{X}}
\def\bx{\mathbf{x}}
\def\bV{\mathbf{V}}
\def\bv{\mathbf{v}}

\def\bU{\mathbf{U}}

\def\bw{\mathbf{w}}
\def\Qtilde{\widetilde{Q}}
\def\Qhat{\widehat{Q}}

\def\Xtilde{\widetilde{X}}

\def\bXtilde{\widetilde{\mathbf{X}}}

\def\fhat{\widehat{f}}
\def\Fhat{\widehat{F}}
\def\fbar{\bar{f}}
\def\ftilde{\widetilde{f}}

\def\btauhat{\widehat{\boldsymbol\tau}}
\def\bSigma{\boldsymbol\Sigma}

\def\Psihat{\widehat{\Psi}}

\def\Psibar{\bar{\Psi}}
\def\sigmahat{\widehat{\sigma}}

\def\etahat{\widehat{\eta}}
\def\etatilde{\widetilde{\eta}}
\def\etastr{\eta^*}
\def\etabar{\bar{\eta}}
\def\etao{\eta^o}
\def\tauhat{\widehat{\tau}}
\def\taubar{\bar{\tau}}
\def\pihat{\widehat{\pi}}
\def\pibar{\bar{\pi}}
\def\muhat{\widehat{\mu}}
\def\mubar{\bar{\mu}}
\def\bgamma{\boldsymbol\gamma}
\def\bgammahat{\widehat{\boldsymbol\gamma}}
\def\bgammabar{\bar{\boldsymbol\gamma}}

\def\omegahat{\widehat{\omega}}

\def\omegabar{\bar{\omega}}

\def\balpha{\boldsymbol\alpha}
\def\balphahat{\widehat{\boldsymbol\alpha}}

\def\balphabar{\bar{\boldsymbol\alpha}}
\def\bbeta{\boldsymbol\beta}
\def\bbetahat{\widehat{\boldsymbol\beta}}

\def\bbetabar{\bar{\boldsymbol\beta}}
\def\kappabar{\bar{\kappa}}

\def\lambdahat{\widehat{\lambda}}
\def\xihat{\widehat{\xi}}

\def\Yone{Y(1)}
\def\Yzer{Y(0)}
\def\Ytee{Y(t)}

\def\bone{\boldsymbol 1}
\def\bzero{\boldsymbol 0}
\def\bI{\boldsymbol I}

\def\Dscr{\mathcal{D}}
\def\Vscr{\mathcal{V}}

\def\E{\mathbb{E}}

\def\P{\mathbb{P}}
\def\R{\mathbb{R}}

\def\indep{\perp\!\!\!\perp}
\def\trans{^{\sf \tiny T}}

\newcommand{\abs}[1]{\left|#1\right|}
\newcommand{\norm}[1]{\left\lVert#1\right\rVert}

\ShortHeadings{Adaptive Combination of Randomized and Observational Data}{Cheng and Cai}
\firstpageno{1}

\begin{document}

\title{Adaptive Combination of Conditional Average 
Treatment Effects Based on Randomized and Observational Data}

\author{\name David Cheng \\
       \addr Biostatistics Center\\
       Massachusetts General Hospital\\
       Boston, MA 02114, USA
       \AND
       \name Tianxi Cai \\
       \addr Department of Biostatistics\\
       Harvard T.H. Chan School of Public Health\\
       Boston, MA 02115, USA}

\editor{December 2021}

\maketitle

\begin{abstract}%
  Data from both a randomized trial and an observational study are sometimes 
  simultaneously available for evaluating the effect of an intervention. The 
  randomized data typically allows for reliable estimation of average treatment 
  effects but may be limited in sample size and patient heterogeneity for 
  estimating conditional average treatment effects for a broad range of patients. 
  Estimates from the observational study can potentially compensate for these 
  limitations, but there may be concerns about whether confounding and treatment 
  effect heterogeneity have been adequately addressed. We propose an approach for 
  combining conditional treatment effect estimators from each source such that it 
  aggressively weights toward the randomized estimator when bias in the observational 
  estimator is detected. This allows the combination to be consistent for a conditional 
  causal effect, regardless of whether assumptions required for consistent estimation 
  in the observational study are satisfied. When the bias is negligible, the estimators 
  from each source are combined for optimal efficiency. We show the problem can 
  be formulated as a penalized least squares problem and consider its asymptotic 
  properties. Simulations demonstrate the robustness and efficiency of the method 
  in finite samples, in scenarios with bias or no bias in the observational estimator. 
  We illustrate the method by estimating the effects of hormone replacement therapy 
  on the risk of developing coronary heart disease in data from the Women's Health 
  Initiative.
\end{abstract}

\section{Introduction}

Randomized clinical trials are regarded as the gold-standard for evaluating 
the causal effect of interventions. Fundamentally, randomization of treatment 
assignment mitigates confounding bias from both measured and unmeasured covariates. 
However, trials are typically designed to evaluate some form of an average 
treatment effect (ATE) of a treatment on an outcome in a specified population. 
To assess treatment effect heterogeneity and inform treatment decisions in practice, 
it is often also of interest to estimate conditional average treatment effects (CATE) 
given a set of observed covariates. Trial data may be undersized for estimation 
and inference about CATEs, as they are powered for testing ATEs \citep{rothwell2005subgroup}. 
Moreover, trials often enroll patient populations that are more narrowly defined 
than those of real-world clinical practice, sometimes raising questions about 
generalizability of findings \citep{cole2010generalizing}. This may also make it 
difficult to estimate CATEs due to the limited heterogeneity in observed patient 
profiles.

Observational studies (OS), in contrast, can potentially collect data for large 
samples from populations that are more representative than trials. 
But in the absence of randomized treatments, 
adjustment for confounding is needed, and there is never a guarantee whether confounding 
has been adequately addressed. These complementary features of trials and OS raise 
the question of whether data from both sources could simultaneously be leveraged 
to yield a more robust and efficient CATE estimator, in situations where it is 
possible to observe parallel data on the same treatments, outcomes, and covariates. 
Availability of such parallel data may 
increase as electronic health records (EHRs) become widely used for clinical 
research and interest grows in embedding pragmatic trials in real-world clinical 
settings \citep{ford2016pragmatic}. Simultaneous analysis of randomized and observational 
data have also drawn recent interest for A/B testing in large-scale online experiments 
\citep{peysakhovich2016combining} and education policy research \citep{kallus2018removing,athey2020combining}.

In this paper, we consider combining CATE estimators from trial and OS data that adapt
to the degree of bias in the OS estimator.  Specificallly, we construct base estimators 
of from trial and OS data that would be consistent for common target estimands 
under working assumptions to address confounding and treatment effect heterogeneity 
that may or may not hold.  The
base estimators are then combined through a linear combination with weights that
minimize an estimate of the the mean square error (MSE).  This leads to a combined
estimator that either minimizes the asymptotic variance when the base estimators
are relatively compatible or aggressively weights toward the trial estimator when
they exhibit sufficient disagreement.  We show the combination is efficient when
the base estimators target a common estimand and smoothly discards the OS estimator 
when it is biased, allowing for incorporation of evidence for OS while minimizing
the risk of introducing confounding bias. 

There have been multiple approaches to integrating randomized and observational
data for estimating treatment effects proposed in the literature.  One broad class 
of methods is to generalize treatment effect estimates from the trial to the OS 
population\citep{cole2010generalizing,kaizar2011estimating,stuart2011use,hartman2015sample,zhang2016new,dahabreh2019generalizing}.  
This typically involves using reweighting, regression, or doubly-robust methods 
to extrapolate estimates of the average treatment effect from the trial to
a target population with a different covariate distribution, using only covariate data from the OS.  
However, the OS data are generally not fully leveraged to separately estimate treatment effects 
and combine with trial estimates.  Methods for synthesizing estimates
from trials and OS have also previously been considered in meta-analysis using aggregate 
data \citep{schmitz2013incorporating,verde2015combining}, which differs from
the framework considered here in which individual-level data are available to inform
the synthesis.

When parallel individual-level data are available, \cite{peysakhovich2016combining} 
proposed using randomized data to estimate a
monotonic transformation of predicted treatment effects based on estimates from
observational panel data, under a strong assumption that the bias in treatment effect 
estimates from observational data preserves unit-level relative rank orderings.
In a similar vein, \cite{kallus2018removing} estimates the CATE in observational
data and calibrates it through a regression of weighted outcomes in the
randomized.  The resulting estimator is consistent for the CATE when 
the bias in OS estimate has a parametric structure, though it is unclear whether
the randomized estimates are incorporated to optimize efficency when the OS estimate
has minimize bias.  Alternatively, \cite{yang2020improved} constructed elegant unbiased
estimating equations for the CATE and a confounding function using a pseudo-outcome
that minimics counterfactual outcomes under control treatment.  Under parametric
specifications for the CATE and confounding function, semiparametric efficiency
theory can be applied to efficiently estimate the CATE and confounding function
by estimating equations.  These approaches generally focus on estimating
the CATE given a full set of pre-treatment covariates.  In contrast, we distinguish
the role of a full set of covariates used to adjust for confounding and a reduced
set of covariates over which to estimate CATE \citep{lee2017doubly,ogburn2015doubly} .  
This allows for more flexibility in estimating the CATE in key covariates.  We
also differ from these other approaches in that we consider a direct weighted
combination of trial and OS estimates, which offers more transparency in synthesis
of estimates from the two data sources.  In a slightly modified data setting, 
\cite{athey2020combining} considered  correcting the bias of treatment effect 
estimates on a long-term outcome in an OS by estimating the bias of a short-term outcome.

Combining randomized and observational treatment effect estimates has previously
been considered in \cite{rosenman2020combining}, assuming that observational estimates
are unconfounded within propensity score (PS) strata.  This was later extended to
allow for biased observational estimates in \cite{rosenman2020combining}, drawing
upon previous work on combining unbiased and biased estimators \cite{green1991james,green2005improved}
to construct shrinkage estimators that minimize unbiased estimates of a weighted average 
of MSEs across strata.  The resulting estimators were shown to achieve lower MSE 
than randomized estimator in finite-sample and asymptotic settings.  While this method
shares similar principles with our proposal, we take a different approach
to constructing the combination.  In particular, we introduce a novel data-driven
approach that takes advantage of linear representations of the trial and OS estimators
to estimate the weights, without explicitly estimating variance nuisance
parameters.  We also allow for estimating the CATE in a reduced set of covariates
without having to discretize into discrete strata and adapt doubly-robust 
(DR) pseudo-outcomes \citep{luedtke2016super,lee2017doubly,kennedy2020optimal} for 
more robust and efficient estimation of the CATE.  

The rest of this paper is organized as follows.  In Section 2 we introduce the observed data, target
CATE estimands, and basic assumptions needed for nonparametric identification of the CATE.
In section 3, we provide an overview of the proposed approach to reduce the bias based
on minimizing the MSE.  We describe the base estimators that are used for combination and 
their linear representation in Section 4 and provide details and theoretic results on the 
combination weights in Section 5.  A simulation study and a data analysis based on parallel 
data from the Women's Health Initiative are given in Sections 6 and 7.  We conclude with
some remarks in Section 8.  We defer proofs to the Appendix.

\section{Framework for Simultaneous Analysis}

Let $Y \in \mathbb{R}$ be an outcome, $T\in \{ 0,1\}$ a treatment, $\bX$ a 
$p$-dimensional vector of baseline covariates, and $Z\in \{ 0,1\}$ an indicator 
of participation in the trial ($Z=0$) or OS ($Z=1$). We assume that $Y$, $T$, and
$\bX$ share common definitions across the trial and OS, though aspects of their joint
distribution may potentially differ.  The trial and OS data consist of independent and 
identicallly distributed (iid) data of the form
$\Dscr_z = \{ (Z_i,\bX_i\trans,T_i,Y_i)\trans: Z_i=z\}$, for $z=0$ and $z=1$ respectively.
The combined data are $\Dscr = \Dscr_0 \cup \Dscr_1$.  Let the size of the each
study be $n_z= \abs{\Dscr_z}$, with the total size as $n=n_0+n_1$.  Let $\bV=g(\bX)$
denote a low-dimensional ($d$-dimensional) reduction of $\bX$ over which we are interested in 
estimating heterogenous treatment effects.  We aim to estimate the CATE over $\bV$ 
for study $Z=z$:
\begin{equation}
  \tau_z(\bv) = \E \left\{\Yone - \Yzer  \mid \bV=\bv,Z=z\right\},
\end{equation}
where $\Ytee$ is the counteractual outcome under treatment $t$.  Here $\tau_z(\bv)$ 
is simply a coarsening of the full CATE $\tau_z(\bx) = \E\left\{ \Yone-\Yzer\mid \bX=\bx,Z=z\right\}$ 
in that $\tau_z(\bv) = \int_{\bx} \tau_z(\bx)dF(\bx\mid\bv,z) $.  We separate the 
roles of $\bV$ from $\bX$ because there is often interest in how treatment effects 
vary in specific set of covariates or other reductions of $\bX$, whereas additional
covariates may be needed to adjust for confounding \citep{ogburn2015doubly,lee2017doubly}.  Additionally, estimation
of treatment effects in $\bX$ may not be feasible unless restrictive parametric 
models are assumed when the dimension of $\bX$ is not small.

Identification of the causal effect $\tau_z(\bv)$ requires assumptions relating 
the distribution of counterfactual variables $\Ytee$ to that of the observed data. 
Our approach is to first assume a baseline set of conditions that hold globally and
then consider data scenarios in which additional assumptions hold for the OS data.

\bigskip
\noindent
{\bf Assumption 1 (Basic conditions)}{\it 
\begin{align}
  &T \indep \left\{ \Yzer,\Yone\right\} \mid Z=0 \label{a:rtrand} \\
  &Y = T\Yone + T\Yzer \text{ almost surely} \label{a:consistency}\\
  &\exists  \epsilon_{\pi}>0 \textrm{ such that } \pi_t(\bx) \in [\epsilon_{\pi},1-\epsilon_{\pi}], \forall \bx,z \textrm{ with } f(\bx,z)>0, \label{a:positivity}\\
  &\exists  \epsilon_{\varrho}>0 \textrm{ such that } \varrho(\bx) \in [\epsilon_{\varrho},1-\epsilon_{\varrho}], \forall \bx \textrm{ with } f(\bx)>0, \label{a:overlap}
\end{align}
where $\pi_t(\bx,z) = \P(T=t\mid\bX=\bx,Z=z)$ and $\varrho(\bx) = \P(Z=z\mid \bX=\bx)$.
}\hfill

\bigskip
\noindent
The unconfoundedness condition \eqref{a:rtrand} distinguishes the trial data from the 
OS data, as treatments in the trial are randomized, and enables the estimation of
a causal effect in trial data.  The consistency
assumption \eqref{a:consistency} links the observed outcomes to counterfactual outcomes.  
The treatment positivity assumption \eqref{a:positivity} stipulates that treatment assignment is not deterministic 
within levels of $\bX$ in both the trial and OS.  These common assumptions are 
foundational and widely adopted in causal inference \citep{kennedy2016semiparametric}.  
The overlapping population assumption \eqref{a:overlap} requires for the distribution 
of $\bX$ in the trial and OS to have a common support. This assumption is used to enable 
reweighting of data from one data source to estimate the CATE in the study population
of the other source.

For the CATE to be identified in the OS data, one may assume that measured 
covariates $\bX$ are sufficiently rich such that conditioning on $\bX$ alleviates
confounding bias.  However, distinct from confounding bias, there may
be differences in the full CATEs $\tau_0(\bx)$ and $\tau_1(\bx)$ induced by 
unmeasured prognostic variables $\bU$ whose conditional distribution given $\bX$ 
differs between the $Z=0$ and $Z=1$ study populations \citep{vanderweele2012confounding}.  For $\tau_0(\bx)$ and 
$\tau_1(\bx)$ to coincide, we may also need to assume that $\bX$ captures all the relevant
prognostic covariates to mitigate such effect modification.  These are encoded in the
following assumptions, which we regard to be optional in that they may or may not hold.

\bigskip
\noindent
{\bf Assumption 2 (No unmeasured confounding or effect modification)}
{\it
\begin{align}
  &T \indep \left\{ \Yone,\Yzer\right\} \mid \bX,Z=1 \label{a:nuca}\\
  &Z \indep \left\{\Yone - \Yzer\right\} \mid \bX \label{a:nuem}
\end{align}
}\hfill

\noindent
If these assumptions hold, it is possible to identify $\tau_0(\bx)$ from the
distribution of both the trial and OS data and subsequently estimate $\tau_z(\bv)$
with the aid of the OS data.  If they are violated, then $\tau_0(\bx)$ is not
identified in the OS data, and we would, in general, prefer to rely on the
trial data to estimate $\tau_z(\bv)$ in case incoporating OS data introduces 
bias.  In practice, these optional assumptions may only hold approximately, though
OS estimates would still be informative of the true $\tau_z(\bv)$.  In this sense,
these assumptions serve as a guide to describe approximate conditions under which
incorporating OS estimates would be reasonable. In the following, we consider a 
combination approach that adaptively combines trial and OS estimators to estimate $\tau_z(\bv)$,
depending on the similarity of estimates from each source.  We then analyze
the estimator in scenarios with and without Assumption 2.

\section{Overview of Approach}
Suppose that we have an estimator $\tauhat_z^r(\bv)$ based on the trial data 
$\Dscr_0$ that is consistent for $\tau_z(\bv)$, for $z=0,1$, under Assumption 1.  
Additionally, suppose we also have an estimator $\tauhat_z^o(\bv)$ based on 
OS data $\Dscr_1$ that is consistent for $\tau_z(\bv)$, under Assumptions 1 and 2.
Consider a linear combination of these base estimators as:
\begin{equation}
  \tauhat_z(\bv;\eta_{\bv}) = \tauhat_z^r(\bv) + \eta_{\bv} \{\tauhat_z^o(\bv)-\tauhat_z^r(\bv)\},
\end{equation}
where $\eta_{\bv}$ is a weight for the OS estimate when estimating the CATE at $\bV=\bv$. 
We suppress the dependence of $\eta$ on $\bv$ in the rest of the paper to ease 
the notation.  In the case where Assumptions 1-2 are satisfied such that $\tauhat_z^r(\bv)$ and $\tauhat_z^o(\bv)$
are both consistent for $\tau_z(\bv)$, we would ideally seek to estimate $\eta$
to minimize their variance.  This leads to the usual precision-weighted style
of estimators.

\begin{lemma}
  Let $\btauhat_z(\bv) = (\tauhat_z^r(\bv), \tauhat_z^o(\bv))\trans$ and 
  $\bw = (1-\eta, \eta)\trans$ such that $\tauhat_z(\bv;\eta) = \bw\trans\btauhat_z(\bv)$.  
  If $r_n\left\{\btauhat_z(\bv)-\tau_z(\bv)\bone\right\} \overset{d}{\to} N(\bzero, \bSigma)$, 
  for some covariance $\bSigma = \big(\begin{smallmatrix}
    \sigma_r^2 & \sigma_{ro}\\
    \sigma_{ro} & \sigma_o^2
  \end{smallmatrix}\big)$ and sequence $r_n>0$, then 
  $r_n \left\{ \tauhat_z(\bv;\eta) - \tau_z(\bv)\right\} \overset{d}{\to}N(\bzero,\bw\trans\bSigma\bw)$.
  The asymptotic variance $\bw\trans\bSigma\bw$ is minimized for
  $\etastr = (\sigma_r^2-2\sigma_{ro})/(\sigma_r^2 + \sigma_o^2 - 2 \sigma_{ro})$.
\end{lemma}

\noindent
Estimates for $\bSigma$ could be used to estimate $\etastr$ as $\etatilde$, which
can be plugged in to obtain an estimator $\tauhat_z(\bv;\etatilde)$ that is efficient, 
provided that $\etatilde$ converges to $\etastr$ at a suitable
rate.  Alternatively, such a precision-weighted estimator $\tauhat_z(\bv;\etatilde)$ can also be motivated
as a solution to minimizing the MSE of $\tauhat_z(\bv;\eta)$ in $\eta$ \citep{lavancier2016general}.

In the case where Assumption 2 fails to hold and $\tauhat_z^o(\bv)$ is biased for $\tau_z(\bv)$, 
the precision-weighted estimator $\tauhat_z(\bv;\etatilde)$ would inherit the bias, as the
weights $\etatilde$ essentially only account for the relative efficiency of the base estimators.  
This prompts whether an \emph{adaptive} weight $\etahat$ can be estimated from the data that either
combines $\tauhat_z^o(\bv)$ with $\tauhat_z^r(\bv)$ for efficiency gain when 
$\tauhat_z^o(\bv)$ is not biased, or is shrunk toward $0$ so as to discard $\tauhat_z^o(\bv)$ when it is 
sufficiently biased.  In brief, we construct such a weight through considering a decomposition of the MSE
for $\tauhat_z(\bv;\eta)$ around the limiting estimand of the trial estimator:  
\begin{equation}
  \E\left\{\tauhat_z(\bv;\eta) - \taubar_z^r(\bv)\right\}^2 = 
  \E\left\{\tauhat_z(\bv;\eta) - \taubar_z(\bv;\eta)\right\}^2 + 
  \eta^2\left\{\taubar_z^o(\bv) -\taubar_z^r(\bv) \right\}^2 + o(r_n), \label{e:mse}
\end{equation}
\noindent
where $\taubar_z(\bv;\eta) = \taubar_z^r(\bv)+ \eta \{\taubar_z^o(\bv)-\taubar_z^r(\bv)\}$, 
$\taubar_z^r(\bv)$ and $\taubar_z^o(\bv)$ are limiting estimands of
$\tauhat_z^r(\bv)$ and $\tauhat_z^o(\bv)$, and $r_n>0$ is some rate.
On the right hand side, the first term is a variance term that is minimized in the
precision weighted estimator $\tauhat_z(\bv;\etatilde)$.  In particular, we show 
that this term can be minimized by performing a least squares regression of 
influence functions from an asymptotic linear expansion of the estimators $\tauhat_z^r(\bv)$ 
and $\tauhat_z^o(\bv)$.  This type of approach using a regression of influence functions
to augment an estimator for efficiency gain has previously been considered for simple
treatment difference estimators in clinical trials \citep{tian2012covariate}.  
The second term is the asymptotic bias of $\tauhat_zz^o(\bv)$ for estimating $\taubar_z^r(\bv)$.
The decomposition thus indicates that weights that minimize the MSE around the trial estimand
$\taubar_z^r(\bv)$ must account for not only the variance but also the bias, which 
can be estimated by $\tauhat_z^o(\bv)-\tauhat_z^r(\bv)$.  

The MSE decomposition in \eqref{e:mse} has the form of a ridge regression objective
function in which the $\ell_2$ penalty on $\eta$ is weighted by 
$\{\tauhat_z^o(\bv)-\tauhat_z^r(\bv)\}^2$.  If the bias were known, such a ridge 
regression approach can be shown to yield an adaptive estimator for $\eta$ in the sense described above.
Unfortunately, when $\tauhat_z^o(\bv)-\tauhat_z^r(\bv)$ is plugged in for the bias,
a direct naive implementation of ridge regression as described above would no longer 
yield adaptive estimates essentially because the bias is being estimated at the same rate as the $\eta$ weights,
which induces bias when $\tauhat_z^o(\bv)$ is not biased (Theorem 6).  Nevertheless,
as discussed in Section 5, that the penalty can be adjusted 
by scaling with a correction factor that shrinks at an appropriate rate to recover 
the adaptive behavior.  Furthermore, in finite samples, even with the correction,
using ridge regression to estimate $\eta$ may still be undesireable because it 
generally does not enable \emph{sparse} solutions for $\eta$ (i.e. for $\eta$ to
be exactly $0$ and completely discard OS).  This is an important feature when combining
evidence from an OS, as there 
may be situations in which $\tauhat_z^o(\bv)$ are so compromised by bias that the 
estimator should not be combined to any degree with $\tauhat_z^r(\bv)$.
To allow for this, we additionally consider a modification of \eqref{e:mse}, 
where a $\ell_1$-penalty is used in place of the $\ell_2$-penalty.

\section{Base Estimators}
We employ nonparametric kernel regression estimators of doubly-robust 
(DR) pseudo-outcomes \citep{luedtke2016super,lee2017doubly,kennedy2020optimal} 
over $\bV$, for base estimators $\tauhat_z^r(\bv)$ and $\tauhat_z^o(\bv)$.  These
pseudo-outcomes are the non-centered efficient influence functions for the average 
treatment effect under a nonparametric model \citep{robins1994estimation}, and
smoothing them in OS data enables doubly-robust estimation so that $\tauhat_z^o(\bv)$ would 
be consistent when either a working parametric model for the PS $\pi_t(\bx,1)$ 
or mean outcomes $\mu_1(\bx,t)$ is correct, under the nonparametric identification
assumptions in Assumption 2.  Smoothing the pseudo-outcome in trial data enables
consistent estimation under the randomization assumption in Assumption 1, which 
generally yield correct specification of working model for $\pi_t(\bx,0)$.  The 
pseudo-outcomes also generally result in more efficient estimators
than those based on PS weighting alone.  In the first stage, the DR pseudo-outcome based on the trial is defined as:

\begin{equation} \label{e:drpseudo}
  \Psihat^r = \frac{T - \pihat_1(\bX,0)}{\pihat_1(\bX,0)\pihat_0(\bX,0)}\left\{ Y - \muhat_0(\bX,T)\right\} + \muhat_0(\bX,1) - \muhat_0(\bX,0),
\end{equation}

\noindent
where $\pihat_t(\bx,z)$ is an estimator for $\pi_t(\bx,z)$ based on
a parametric working model fit in study $Z=z$, and 
$\pihat_t(\bx,z) = \pi_t(\bx,z;\balphahat_z)$ for some estimated parameters $\balphahat_z$ 
(e.g. logistic regression of $T\sim\bX$).  Similarly, $\muhat_z(\bx,t)$ is an estimator 
for $\mu_z(\bx,t)= \E(Y\mid \bX=\bx,T=t,Z=z)$ based on fitting a parametric working model
fit in study $Z=z$, and $\muhat_z(\bx,t)=\mu_z(\bx,t;\bbetahat)$ for some $\bbetahat_z$ (e.g. 
linear regression of $Y \sim \bX + T$).  The working parametric models are, in general, 
approximations that could possibly be misspecified. An exception is that under \eqref{a:rtrand}
in Assumption 1, the PS working model in the trial would be correctly 
specified due to randomization.  The DR pseudo-outcome based on the 
OS $\Psi^o(\bv)$ is defined analogously, except that it replaces
$\pihat_t(\bX,1)$ for $\pihat_t(\bX,0)$ and $\muhat_1(\bX,t)$ for $\muhat_0(\bX,t)$
in \eqref{e:drpseudo}, for $t=0,1$.  

In the second stage, the kernel regression regresses $\Psihat^r$ and $\Psihat^o$ 
over $\bV$ to yield the trial and OS base estimators:

\begin{align}
  &\tauhat_z^r(\bv) = n^{-1}\sum_{j=1}^n K_h(\bV_j -\bv)I(Z_j=0) \omegahat(\bX_j)^{z}\Psihat_j^r /  \fhat_z^r(\bv)\\
  &\tauhat_z^o(\bv) = n^{-1}\sum_{j=1}^n K_h(\bV_j -\bv)I(Z_j=1) \omegahat(\bX_j)^{z-1}\Psihat_j^o /  \fhat_z^o(\bv),
\end{align}

\noindent
where $K_h(\cdot) = K(\cdot/h)/h^d$ for some $d$-dimensional multivariate kernel $K(\cdot)$, $\omegahat(\bx)$
is an estimator of the odds of participating in the OS vs. trial $\omega(\bx) = \P(Z = 1\mid \bX=\bx) / \P(Z = 0\mid \bX=\bx)$, 
and $\fhat_z^r(\bv) = n^{-1}\sum_{j=1}^n K_h(\bV_j-\bv)I(Z_j=0)\omegahat(\bX_j)^z$
is the density estimate, with $\fhat_z^o(\bv)$ being defined analogously based on OS data.  
The estimate of OS odds $\omegahat(\bx)$ are obtained by fitting a 
logistic regression or some other parametric working model for $Z\sim \bX$ in 
the pooled data $\Dscr$ and $\omegahat(\bx) =\omega(\bx;\bgammahat)$ for some estimated 
parameters $\bgammahat$.  These odds weights $\omegahat(\bx)$ further reweight
observations to reflect the target $Z=z$ population based on the distribution of
covariates $\bX$ \citep{westreich2017transportability}.  The odds weights can also
be omitted if the interest is in estimating the CATE for the pooled population consisting
of observations from both the trial and OS.
If covariate data on $\bX$ for another target population outside of the trial or 
OS is available, that could be used to estimate the odds of selection into 
another target population relative to either the trial or OS population at hand, which 
can also be used to reweight estimates to target estimation of CATEs in other
populations.  In this paper, we focus on the CATE in the $Z=0$ and $Z=1$ populations.

Locally constant kernel regression is adopted for the second stage, but other methods 
such as locally linear regression can also be considered.  The
key feature we require of the base estimators is that they admit an asymptotic linear
expansion in which the centered and appropriately scaled estimators 
can be decomposed as a sum of independent terms and a higher-order remainder, which 
enables the minimization of the MSE in \eqref{e:mse} to be approximated by
a penalized least squares criterion, in terms of the constituent influence functions.
The following result identifies such a linear representation for locally constant
estimators.

\begin{lemma}
Let $\pibar_t(\bx,z)$, $\mubar_z(\bx,t)$, $\omegabar(\bx)$, $\fbar_z^r(\bv)$, and 
$\fbar_z^o(\bv)$  denote asymptotic limits of $\pihat_t(\bx,z)$, $\muhat_z(\bx,t)$, 
$\omegahat(\bx)$, $\fhat_z^r(\bv)$, and $\fhat_z^o(\bv)$, respectively, with $\Psibar_i^r$ 
and $\Psibar_i^o$ defined similarly as $\Psihat_i^r$ and $\Psihat_i^o$, except 
replacing the nuisance functions $\pihat_t(\bx,z)$ and $\muhat_z(\bx,t)$ by 
$\pibar_t(\bx,z)$ and $\mubar_z(\bx,t)$. Suppose that $\norm{\balphahat_z-\balphabar_z}=O_p(n^{-1/2})$ and 
$\norm{\bbetahat_z-\bbetabar_z} = O_p(n^{-1/2})$, for $z=0,1$, and
$\norm{\bgammahat-\bgammabar}=O_p(n^{-1/2})$, where $\balphabar_z,\bbetabar_z,\bgammabar$
are the corresponding limits regardless of working parametric models are correctly specified.  
The base estimators $\tauhat_z^r(\bv)$ and $\tauhat_z^o(\bv)$ admit asymptotic linear expansions:
\begin{align*}
  &(nh^d)^{1/2}\left\{ \tauhat_z^r(\bv) - \taubar_z^r(\bv) \right\} = (nh^d)^{-1/2}\sum_{i=1}^n \xi_{z,i}^r(\bv) + o_p(1) \\
  &(nh^d)^{1/2}\left\{ \tauhat_z^o(\bv) - \taubar_z^o(\bv) \right\} = (nh^d)^{-1/2}\sum_{i=1}^n \xi_{z,i}^o(\bv) + o_p(1),
\end{align*}
where $\xi_{z,i}^r(\bv) = K\{(\bV_i -\bv)/h\}I(Z_i= 0)\omegabar(\bX_i)^{z}\left\{\Psibar_i^r - \taubar_z^r(\bv) \right\}/\fbar_z^r(\bv)$  and
$\xi_{z,i}^o(\bv) = K\{(\bV_i -\bv)/h\}I(Z_i= 1)\omegabar(\bX_i)^{z-1}\left\{\Psibar_i^o - \taubar_z^o(\bv) \right\}/\fbar_z^o(\bv)$. 
The linear terms $\xi_{z,i}^r(\bv)$ are iid and have conditional mean $0$
such that $\E\left\{ \xi_{z,i}^r(\bv) \mid \bV=\bv\right\} = 0$, and likewise for
$\xi_{z,i}^o(\bv)$.
\end{lemma}

\noindent
We refer to $\xi_{z,i}^r(\bv)$ and $\xi_{z,i}^o(\bv)$ 
as the \emph{influence functions} for $\taubar_z^r(\bv)$ and $\taubar_z^o(\bv)$.

This representation indicates that $\tauhat_z^r(\bv)$ and $\tauhat_z^o(\bv)$ are consistent for
estimands $\taubar_z^r(\bv)$ and $\taubar_z^o(\bv)$ in general, but further explanation
is needed to clarify how $\taubar_z^r(\bv)$ and $\taubar_z^o(\bv)$ relate to the 
CATE $\tau_z(\bv)$.  As the trial estimator $\tauhat_z^r(\bv)$ relies on data with
randomized treatments, its estimand $\taubar_z^r(\bv)$ can be shown to either coincide with
or approximate $\tau_z(\bv)$ under basic assumptions in Assumption 1.  It may not 
exactly coincide with $\tau_z(\bv)$ when targeting the OS population because the 
odds weight are estimated by $\omegahat(\bx)$, which may be inconsistent (i.e. such 
that $\omegabar(\bx) \neq \omega(\bx)$) under a misspecified working model for study 
participation.  We formalize the assumption that this working model is correctly 
specified in the following.

\bigskip
\noindent
{\bf Assumption 3 (Correct specification of OS participation working model)}{\it

\noindent
Let $\varrho(\bx)=\P(Z=1\mid\bX=\bx)$ be the probability of participating in the OS
among those with $\bX=\bx$.  We assume that $\varrho(\bx)$ is correctly specified
by a working parametric model $\varrho(\bx;\bgamma)$ such that  
$\varrho(\bx) \in \{ \varrho(\bx;\bgamma) : \bgamma \in \R^{p_{\varrho}}\}$, for some positive integer $p_{\varrho}$.
}\hfill

\bigskip
\noindent
Under this assumption, the odds of OS participation $\omega(\bx)$ can be consistently
estimated as $\omegahat(\bx) = \varrho(\bx;\bgammahat)/\{1-\varrho(\bx;\bgammahat)\}$.  
A special case in which such a working model would hold is when the distribution 
of $\bX$ are the same between the two populations, for example. 

We now identify specific scenarios in which the trial estimand coincides with 
the target estimand such that $\taubar_z^r(\bv) = \tau_z(\bv)$.  As the adaptive 
combination will be constructed to be consistent for the trial estimand 
$\taubar_z^r(\bv)$, these scenarios in turn identify when the adaptive combination 
would be consistent for $\tau_z(\bv)$ or an approximation of $\tau_z(\bv)$.

\begin{lemma} \label{l:taubarr}
  Under \eqref{a:rtrand}-\eqref{a:positivity} in Assumption 1, $\taubar_0^r(\bv)=\tau_0(\bv)$.
  Under Assumption 1, then $\taubar_1^r(\bv) = \E_{\bXtilde\mid \bV}\left\{\tau_1(\bX) \mid \bV=\bv\right\}$,
  where $\E_{\bXtilde\mid\bV}(\cdot \mid \bV=\bv)$ denotes a conditional expectation
  for $\bX$, characterized by the tilted density $\ftilde(\bx\mid\bV=\bv) = \P(Z=0\mid\bX=\bx)\omega(\bx;\bgammabar)f(\bx\mid\bV=\bv)/\kappabar(\bv)$,
  where $\kappabar(\bv) = \int \P(Z=0\mid\bX=\bx)\omega(\bx;\bgammabar)f(\bx\mid\bV=\bv)d\bx$.
  Under Assumption 1, the no unmeasured effect modification condition \eqref{a:nuem} in 
  Assumption 2, and correctly specified models for OS participation in Assumption 3, 
  $\taubar_1^r(\bv)=\tau_1(\bv)$.  
\end{lemma}

\noindent
Based on this, when targeting the trial population, 
$\taubar_0^r(\bv)=\tau_0(\bv)$ under only the basic assumptions of Assumption 1.
When targeting the OS population, $\taubar_1^r(\bv)=\tau_1(\bv)$ if there are no 
unmeasured effect modifiers and the model for OS participation is
correctly specified.  Even if \emph{only} Assumption 1 held, $\taubar_1^r(\bv)$
is still a meaningful estimand, in that it is a reweighted version of the CATE in 
the OS population and represents a reweighted version of a conditional causal effect of treatment. 

The OS estimator $\tauhat_z^o(\bv)$ estimates $\tau_z(\bv)$ through adjustment for
$\bX$ in working parametric models for the propensity score $\pi_t(\bx,1)$ and 
mean outcomes $\mu_1(\bx,t)$.  As these models may also be misspecified, we formalize
the condition that at least one of these models is correctly specified in the following
assumption.

\bigskip
\noindent
{\bf Assumption 4 (Correct specification of PS or outcome working models)}{\it

\noindent
Either the PS in the OS data is correctly specified by a parametric model 
$\pi_t(\bx,1;\balpha)$ or the mean outcomes are corectly specified by $\mu_1(\bx,t;\bbeta)$ 
such that $\pi_t(\bx,1) \in \{ \pi_t(\bx,1;\balpha):\balpha\in\R^{p_{\pi}}\}$ or $\mu_1(\bx,t;\bbeta)\in \{ \mu_1(\bx,t;\bbeta): \bbeta\in\R^{p_{\mu}}\}$, but not necessarily both, for 
some positive integers $p_{\pi}$ and $p_{\mu}$.
}\hfill

\bigskip
We now identify scenarios when $\taubar_z^o(\bv)=\tau_z(\bv)$, in which case 
the OS estimates are informative for the CATE and can be combined with the trial estimates for efficiency.

\begin{lemma}
Under \eqref{a:consistency}-\eqref{a:positivity} in Assumption 1, \eqref{a:nuca}
in Assumption 2, and Assumption 4, $\taubar_1^o(\bv)=\tau_1(\bv)$.  Under these assumptions
and \eqref{a:overlap} in Assumption 1 and Assumption 3, we also have that $\taubar_0^o(\bv)=\tau_0(\bv)$.
\end{lemma}

\noindent
Taken together with Lemma \ref{l:taubarr}, when targeting the trial population,
$\tauhat_0^r(\bv)$ is consistent for $\tau_0(\bv)$ under Assumption 1, and $\tauhat_z^o(\bv)$
is also consistent for $\tau_0(\bv)$ additionally under Assumption 2-4.  When targeting the OS population,
$\tauhat_1^r(\bv)$ is consistent for $\tau_1(\bv)$ under Assumption 1, \eqref{a:nuem} 
in Assumption 2, and Assumption 3.  We also have that $\tauhat_1^o(\bv)$ is consistent
for $\tau_1(\bv)$ additionally under \eqref{a:nuca} in Assumption 2, and Assumption 4.

\section{Adaptive Combination}
To obtain adaptive estimates of $\eta$, we return to the MSE in \eqref{e:mse}.
Using the asymptotic linear expansions, we obtain that the MSE can be approximated
by a least squares criterion plus an additional term that accounts for the bias in the OS estimate.

\begin{lemma}
  Under regularity conditions specified in the Appendix, when locally constant nonparametric kernel regression 
  are used for the base estimators $\tauhat_z^r(\bv)$ and $\tauhat_z^o(\bv)$, the
  MSE around $\taubar_z^r(\bv)$ can be expressed as:
  \begin{align*}
    \E\left\{\tauhat_z(\bv;\eta) - \taubar_z^r(\bv)\right\}^2 &= 
    n^{-1}h^{-2d}\E\left[\xi_{z,i}^r(\bv) - \eta\left\{\xi_{z,i}^r(\bv) - \xi_{z,i}^o(\bv)\right\} \right]^2 \\
    &\qquad  + \eta^2\left\{\taubar_z^o(\bv) -\taubar_z^r(\bv) \right\}^2  +o\{(nh^d)^{-1/2}\}.
  \end{align*}
\end{lemma}

\noindent
This representation of the MSE suggests that $\eta$ can be estimated by a scaled 
empirical estimate:

\begin{align}
  \Qtilde_z(\eta; \bv) &= \sum_{i=1}^n \left[\xihat_{z,i}^r(\bv) - \eta\left\{\xihat_{z,i}^r(\bv) - \xihat_{z,i}^o(\bv;)\right\}\right]^2
  + n^{2(1-\beta)}h^{2d}\eta^2\left\{\tauhat_z^o(\bv) -\tauhat_z^r(\bv) \right\}^2 \label{e:Qridge} \\
  &= \sum_{i=1}^n \left( \xihat_{z,i}^r(\bv) - \eta \left[ \xihat_{z,i}^r(\bv)-\xihat_{z,i}^o(\bv)-n^{1-\beta}h^d\left\{\tauhat_z^o(\bv)-\tauhat_z^r(\bv)\right\}\right]\right)^2 \label{e:QLS}
\end{align}
where $\xihat_{z,i}^r(\bv)= K\{(\bV_i -\bv)/h\}I(Z_i= 0)\omegahat(\bX_i)^z\left\{\Psihat_i^r - \tauhat_z^r(\bv) \right\}/\fhat_z^r(\bv)$ 
is the influence function in which the unknown parameters are subsituted by their estimates
and $\xi_{z,i}^o(\bv)$ is analogously defined.  The $n^{-\beta}$ term
is an additional correction on the estimated bias.  It can be shown that when 
$\beta=0$ and there is no correction, the resulting estimator of $\eta$ from minimizing 
$\Qtilde_z(\eta;\bv)$ fails to adapt to biased OS data in that it 
may not be consistent for $0$ when $\taubar_z^o(\bv)\neq \taubar_z^r(\bv)$.  In 
this case, the adaptive combination can be severely biased because it potentially
allows for combination OS estimators even when it is not consistent for the same
estimand as the trial estimator.  Such nettlesome behavior is one consequence 
of relying on estimated biases. Nevertheless, this can 
potentially be avoided by scaling the bias penalty term with a correction factor 
$n^{-2\beta}$, for an appropriate $\beta>0$ given below. 

The empirical criterion $\Qtilde_z(\eta;\bv)$ coincides with either that of a ridge 
regression of $\xihat_{z,i}^r(\bv)$ 
on $\xihat_{z,i}^r(\bv)-\xihat_{z,i}^o(\bv)$ that weights the 
$\ell_2$ penalty term by the squared bias scaled by $n^{2(1-\beta)}h^{2d}$, or directly by a least squares of
$\xihat_{z,i}^r(\bv)$ on 
$\xihat_{z,i}^r(\bv)-\xihat_{z,i}^o(\bv)-n^{1-\beta}h^d\left\{\tauhat_z^o(\bv)-\tauhat_z^r(\bv)\right\}$,
using that $\sum_{i=1}^n \xihat_{z,i}^r(\bv) =\sum_{i=1}^n \xihat_{z,i}^o(\bv)=0$.
In the following, we show that when $\eta$ is estimated by minimizing $\Qtilde_z(\eta;\bv)$,
we obtain an adaptive estimator in that it is consistent for the optimal
MSE weights if the OS estimates are unbiased in that $\taubar_z^r(\bv)=\taubar_z^o(\bv)$.  
Otherwise, the estimate of $\eta$ is consistent for $0$ and discards the OS estimate
in large samples.

\begin{theorem}
Let $\beta \in (0,\frac{1-\alpha}{2})$.  The minimizier $\etatilde = argmin_{\eta}\Qtilde(\eta;\bv)$ 
is an adaptive weight in that $\etatilde = O_p(n^{\alpha+2\beta-1})$ when $\taubar_z^r(\bv)\neq \taubar_z^o(\bv)$
and $\etatilde - \etastr = O_p\left\{(nh)^{-1/2}\right\}$ when $\taubar_z^r(\bv) = \taubar_z^o(\bv)$, where
$\etastr$ is the minimizer of the asymptotic variance of $\tauhat_z(\bv;\eta)$ when $\taubar_z^r(\bv)=\taubar_z^o(\bv)$.
\end{theorem}

The resulting estimator $\tauhat(\bv;\etatilde)$ is thus consistent for $\taubar_z^r(\bv)$
regardless of the bias of $\tauhat_z^o(\bv)$.  The correction order $\beta$ controls
the degree to which weights are similar to $\etastr$ or are shrunken to $0$.  A 
larger $\beta$ leads to a greater discounting of the bias term in $\Qtilde_z(\eta;\bv)$,
and the procedure aims to more directly estimate weights that minimize the asymptotic variance
assuming no bias.  But if there is bias, a larger $\beta$ also slows the convergence
of $\etatilde$ to $0$.  In practice, we found that manually choosing appropriate value for $\beta$ can be
challening, and faulty choices can result in poor performance of $\tauhat(\bv;\etatilde)$.
Minimizing $\Qtilde_z(\eta;\bv)$ also generally does not yield weights that are exactly 0,
which can still lead to combinations with poor performance in finite samples when the
OS estimator $\tauhat_z^o(\bv)$ is severely biased.  As an alternative, we considered replacing the 
weighted $\ell_2$ penalty in \eqref{e:Qridge} with a weighted $\ell_1$ penalty:

\begin{equation}
  \Qhat_z(\eta; \bv) = \sum_{i=1}^n \left[\xihat_{z,i}^r(\bv) - \eta\left\{\xihat_{z,i}^r(\bv) - \xihat_{z,i}^o(\bv)\right\}\right]^2
  + \lambda_n\abs{\eta}\left\{\tauhat_z^o(\bv) -\tauhat_z^r(\bv) \right\}^2, \label{e:Qlasso}
\end{equation}

\noindent
where $\lambda_n$ is some tuning parameter.  This turns the problem into an adaptive
LASSO \cite{zou2006adaptive} problem in which the penalty on $\eta$ is large when
the estimated bias is large and vice versa when the estimated bias is small. By
using the $\ell_1$ penalty, the procedure may estimate $\eta$ to be exactly 0 so
as to discard the OS estimate $\tauhat_z^o(\bv)$ when it is severely biased.  If 
$\tauhat_z^o(\bv)$ is not severely biased, the first term in $\Qhat_z(\eta;\bv)$
would dominate and lead to an estimator that still optimizes the asymptotic variance, as 
in \eqref{e:Qridge}.  We also consider a data-driven procedure to estimate suitable 
values of the tuning parameter $\lambda_n$. The resulting estimator of $\eta$ is 
also adaptive with a suitable choice of the tuning parameter.

\begin{theorem}
  Let $\lambda_n = O(n^{\gamma})$, where $\gamma \in (1-\alpha, 2(1-\alpha))$.  Then 
  $\etahat = argmin_{\eta}\Qhat_z(\eta;\bv)$ is an adaptive weight in that $\P(\etahat= 0 ) \to 1$
  when $\taubar_z^r(\bv)\neq \taubar_z^o(\bv)$ and $\etahat - \etastr = O_p\left\{(nh^d)^{-1/2}\right\}$
  when $\taubar_z^r(\bv)= \taubar_z^o(\bv)$.
\end{theorem}

The resulting estimator $\tauhat(\bv;\etahat)$ is consistent for $\taubar_z^r(\bv)$
regardless of the bias of the OS estimator $\tauhat_z^o(\bv)$ and discards $\tauhat_z^o(\bv)$ by 
setting $\etahat=0$ when severe bias is detected.

\subsection{Inference for Adaptive Combination $\tauhat_z(\bv)$}

We consider estimating standard errors (SE) for $\tauhat_z(\bv)$ directly using
the influence functions for $\tauhat_z^r(\bv)$ and $\tauhat_z^o(\bv)$. Based on
the expansions in Lemma 2, the combination can also be expanded as:
\begin{align*}
  (nh^d)^{1/2}\left\{\tauhat_z(\bv)-\taubar_z^r(\bv)\right\} &= (nh^d)^{-1/2}\sum_{i=1}^n \xi_{z,i}^r(\bv) + \etabar \left\{\xi_{z,i}^o(\bv)-\xi_{z,i}^r(\bv)\right\}\\
  &\qquad + \etabar\left\{\taubar_z^o(\bv)-\taubar_z^r(\bv)\right\} + (\etahat - \etabar) \left\{\taubar_z^o(\bv)-\taubar_z^r(\bv)\right\}+ o_p\left\{(nh^d)^{-1/2}\right\},
\end{align*}
where $\etabar$ is the limiting value of $\etahat$ in general.  On the right hand side,
the the terms involving the bias of the OS estimator can be considered to be negligible
in large samples.  This follows if $\taubar_z^r(\bv)=\taubar_z^o(\bv)$.  Otherwise,
if $\taubar_z^r(\bv)\neq\taubar_z^o(\bv)$, then $\etabar=0$ and $\P(\etahat=0)\to 1$,
which also leads the additional terms involving the bias to be negligible.  It can
be shown through Liapunov's central limit theorem that $(nh^d)^{1/2}\left\{ \tauhat_z(\bv)-\taubar_z^r(\bv)\right\}\overset{d}{\to}N(0,\sigma^2(\bv))$,
where $\sigma^2(\bv) = h^{-1}\E\left\{\etabar\xi_{z,i}^o(\bv) + (1-\etabar)\xi_{z,i}^r(\bv)\right\}^2$.
We estimate the standard error of $\tauhat(\bv)$ as $\left\{\sigmahat^2(\bv)/(nh^d)\right\}^{1/2}$, 
where: 
\begin{equation}
  \sigmahat_z^2(\bv)=(nh^d)^{-1}\sum_{i=1}^n \left\{\etahat\xihat_{z,i}^o(\bv) + (1-\etahat)\xihat_{z,i}^r(\bv)\right\}^2.  
\end{equation}
In scenarios where the estimated bias is appreciable and $\etahat \neq 0$,
\begin{equation}
  \sigmahat_{z,c}^2(\bv)=(nh^d)^{-1}\sum_{i=1}^n \left\{\etahat\xihat_{z,i}^o(\bv) + (1-\etahat)\xihat_{z,i}^r(\bv)\right\}^2 + \etahat^2\left\{\tauhat_z^o(\bv)-\tauhat_z^r(\bv)\right\}^2
\end{equation}
can also be used as a more conservative alternative.  Given SE estimates, pointwise
confidence intervals for $\tauhat_z(\bv)$ can be constructed based on normal 
approximation.  We find in simulations that the correction with the estimated bias 
can achieve better performance in terms of coverage in finite samples.

\subsection{Choice of Tuning Parameter $\lambda_n$}

To select a suitable $\lambda_n$, let $\Dscr_{v}=\left\{(Z_i,\bX_i\trans,T_i,Y_i)\trans\right\}$
denote an additional validation dataset consisting of pooled randomized and observational
data with the same distribution as that of $\Dscr$.  In practice, if the available 
dataset is large enough, the entire pooled data can initially be partioned into $\Dscr$ and $\Vscr$.
We proceed in the following with the understanding that, in the context of smaller datasets, the procedure can be repeated
switching the role of the training dataset $\Dscr$ and validation dataset $\Vscr$
to gain efficiency through cross-fitting (for example, as done in \cite{robins2008higher,chernozhukov2018double}) or other sample-splitting techniques. 
We select $\lambda_n$ by minimizing an out-of-sample estimate of the mean integrated 
square error (MISE) in the validation data $\Dscr_{v}$:

\begin{equation*}
  R_z(\lambda) = \sum_{\bX_i \in \Dscr_{v}}\left\{\tauhat_z(\bV_i;\lambda)-\tauhat_z^{r,\Dscr_v}(\bV_i)\right\}^2,
\end{equation*}

\noindent
where $\bV_i = g(\bX_i)$, $\tauhat_z(\bv;\lambda)$ is the adaptive combination $\tauhat_z(\bv)$
for a given tuning parameter $\lambda$ estimated in the training data $\Dscr$, and
$\tauhat_z^{r,\Dscr_v}(\bv)$ is the trial estimator estimated from the validation data $\Dscr_v$.
Plugging in a consistent estimator from an independent dataset for the unknown parameter
in a MSE criterion has previously been shown to be a reasonable approach for selecting
among estimators to minimize the true MSE \citep{brookhart2006semiparametric}.
As we are using an integrated criteria across $\bV$ to select the tuning parameter, 
we do not consider separate tuning parameters for estimates at
different $\bv$.  This unified approach effectively allows for borrowing
of information across $\bv$ when estimate separate weights $\eta = \eta_{\bv}$ across $\bv$.
We also restrict the candidate $\lambda$ to yield a unified decision across $\bv$
as to whether the adaptive combination should allow for combination or not.
Specifically, we consider the union of a grid for $\lambda$ of evenly spaced 
values on the log scale within the interval $[\epsilon\lambda_{max},\lambda_{max}]$ with $\{\lambda_{max}^+\}$
for the set of candidate $\lambda$'s to evaluate in the validation procedure, 
where $\lambda_{max}=\text{min}_{\bv\in \Vscr}\left[\etao_{\bv}/\{\tauhat_z^r(\bv)-\tauhat_z^o(\bv)\}^2\right]$,
$\lambda_{max}^+=\text{max}_{\bv\in \Vscr}\left[\etao_{\bv}/\{\tauhat_z^r(\bv)-\tauhat_z^o(\bv)\}^2\right]$,
$\eta_{\bv}^o$ is the least square estimator regressing $\xihat_{z,i}^r(\bv) \sim \left\{\xihat_{z,i}^o(\bv)-\xihat_{z,i}^r(\bv)\right\}$ for a given $\bv$, 
$\Vscr$ is the set of evaluation points of $\bv$ for which we want to estimate 
$\tau_z(\bv)$, and $\epsilon>0$ is some small constant.  This set of candidate values 
for $\lambda$ effectively allows
for a range of $\eta$ across $\bv$ ranging from the least square estimate $\etao_{\bv}$
to $0$.  It excludes the possibility for $\eta$ to be exactly $0$ at some evaluation points $\bv\in\Vscr$
and non-zero at other points $\bv\in\Vscr$, as bias in $\tauhat_z(\bv)$ at any $\bv$ may cast suspicion
of its reliability at other $\bv$.  

\section{Simulation Studies}

We evaluated the bias, root mean square error (RMSE), and coverage 
of the proposed estimator $\tauhat_z(\bv)$ relative to the trial estimator 
$\tauhat_z^r(\bv)$ and the OS estimator $\tauhat_z^o(\bv)$ in simulations to
assess their finite-sample performance.  Each of the estimators are estimated in 
training data $\Dscr$ of size $n$, and we assume that a large external dataset
$\Dscr_v$ of size $n_v=20,000$ is available for selecting the tuning parameter 
$\lambda_n$.  A large validation set is needed accumulate 
sufficient number of trial participants to achieve satisfactory performance for 
tuning parameter selection, as we assume that only a small fraction of the combined data
are from the trial.  We considered both the pointwise performance
at different values of $\bv$ and also the integrated performance across $\bv$.  For example,
for the RMSE, we estimated:
\begin{align*}
  &RMSE(\bv) = \left[R^{-1}\sum_{r=1}^R\left\{\tauhat_z^{(r)}(\bv) -\tau_z(\bv)\right\}^2\right]^{1/2} \\
  &RMSE = \left[(nR)^{-1}\sum_{r=1}^R\sum_{i=1}^n\left\{\tauhat_z^{(r)}(\bV_i) -\tau_z(\bV_i)\right\}^2\right]^{1/2},
\end{align*}
where $\tauhat_z^{(r)}(\bv)$ denotes the adaptive estimator estimated from the
$r$-th simulation replication, for $r=1,\ldots,R$.  We considered data scenarios
in which the working parametric model for OS participation $\varrho(\bx;\bgamma)$
is correctly specified and models for the PS and mean outcomes in the
OS, $\pi_1(\bx,1;\balpha)$ and $\mu_1(\bx,t;\bbeta)$ are either both correct or both misspecified.

\subsection{Simulated Parallel Trial and OS Data}

The data are simulated based on the example in \cite{kang2007demystifying}.
We assume that there are $p=4$ baseline covariates and that $\bX_i \overset{iid}{\sim} N(\bzero, \bI)$.  
The trial selection indicators are assumed to be $Z_i \overset{iid}{\sim} Ber\{\varrho(\bX_i)\}$, 
where true OS selection probabilities are:
\begin{equation} \label{e:simstudyselect}
  \varrho(\bx) = expit\left\{(1,\bx\trans)\bgamma\right\},
\end{equation}
with $\bgamma = (2.5,.1\bone_4)$.  This trial selection mechanism selects 
a minority of individuals from the population into the trial, around 
8\%.  There are also some weak study selection effects such that the distribution of $\bX$ is shifted towards
larger values in the OS.  The treatment statuses are generated as 
$T_i \overset{iid}{\sim} Ber\{\pi_1(\bX_i,Z_i)\}$, where the true PS are:
\begin{equation} \label{e:simps}
  \pi_1(\bx, z) = expit\left\{(1,\bx\trans,z,z\bx\trans)\balpha\right\},
\end{equation}
with $\balpha = (0,\bzero_4,0,-1,.5,-.25,-.1)$.  This allows treatment to be
randomized 1:1 for trial participants and to be selected based on covariates $\bX$
for OS participants, with the strength of selection varying among the covariates.
We consider continuous outcomes generated by $Y_i = \mu_{Z_i}(\bX_i,T_i) +\varepsilon_{i}$, 
where $\varepsilon_i \overset{iid}{\sim}N(0,1)$ and the mean outcomes are given by:
\begin{equation} \label{e:simoutcome}
  \mu_z(\bx,t) = (1,\bx\trans)\bbeta,
\end{equation}
with $\bbeta = (210,27.4,13.7 \bone_3)$.  The mean outcomes are assumed to be
structurally independent of treatment $T$ and study selection $Z$ such that the
true CATE given $\bX=\bx$ in either study population is $\tau_0(\bx)=\tau_1(\bx)=0$, for all $\bx$.
Under this setup, we have that Assumption 2 is satisfied so that there is no unmeasured
confounding and effect modification given $\bX$.

In the correctly specified scenario, we assume that $\bX_i$'s are directly observed and
use the specifications in \eqref{e:simstudyselect}, \eqref{e:simps}, and \eqref{e:simoutcome}
for the working parametric models when constructing the pseudo-outcomes $\Psihat^r$ and $\Psihat^o$.  
In misspecified working models scenario, we assume that a transformed 
version of the covariates $\bXtilde_i$ is observed, where $\Xtilde_{i1} = exp(X_{i1}/2)$,
$\Xtilde_{i2} = X_{i2}/(1+exp(X_{i1})) + 10$, $\Xtilde_{i3} = (X_{i1}X_{i3}/25 + 0.6)^3$, 
and $\Xtilde_{i4} = (X_{i2}+X_{i4}+20)^2$.  We then assume the specifications
in \eqref{e:simstudyselect}, \eqref{e:simps}, and \eqref{e:simoutcome} for the 
working models, using $\bXtilde$ rather than $\bX$.  This implies that the true 
models are highly non-linear in $\bXtilde$ and results in misspecification of both 
the outcome and PS working models.  The trial selection model is also misspecified, 
but the impact on estimating the true CATE is expected to be minimal, as the true 
CATE do not vary over $\bX=\bx$.

We focused specifically on estimating $\tau_1(\bv)$, the CATE in the OS population 
given $\bV=\Fhat_{X_1}^{\Dscr_v}(X_{1})$, where $\Fhat_{X_1}^{\Dscr_v}(\cdot)$ 
denotes  the empirical cumulative distribution function of $X_1$ estimated in the validation 
set $\Dscr_{v}$.  Even though $\bV$ is based on an estimated transformation $\Fhat_{X_1}^{\Dscr_v}(\cdot)$, we 
find that there is minimal impact on inferences and regard it as a fixed
transformation of $\bX$.  We implement the trial, OS, and adptive estimators $\tauhat_1^r(\bv)$, 
$\tauhat_1^o(\bv)$, and $\tauhat_1(\bv)$, identifying an optimal tuning parameter for $\tauhat_1(\bv)$
in the separate validation set $\Dscr_v$.  For pointwise evaluations, we used an 
evenly-spaced grid for $\bv$ over the interval $(0,1)$.  To assess the pointwise performance 
of CIs for $\tauhat_1(\bv)$, we implemented the the 
SE estimate  based on $\sigmahat_z^2(\bv)$ as well as the conservative 
alternative $\sigmahat_{z,c}^2(\bv)$ as proposed Section 5.1.  

\subsection{Simulation Results}

Table 1 presents the bias and RMSE.  Under correctly specified models, both the trial
and OS estimators are unbiased, with the OS estimator being substantially more
efficient.  Despite imbalances in prognostic covariates between treatment groups
in the OS data, the OS estimator remains free of bias through adjustment with the
DR pseudo-outcomes.  The adaptive estimator also maintains minimal bias and achieves
RMSE similar to that of OS and is substantially more efficient, having mean integrated
square error that is 83-88\% lower than the trial estimator.  Under severe misspecification of
all working models, the OS estimator incurs high bias that is sustained even with larger 
training samples. The trial estimator general has low bias relative to RMSE.  The 
adaptive estimator has substantially lower bias than OS that decreases with larger 
training samples. In particular, there is relatively minimal bias relative to RMSE 
with a training set size of $n=10,000$, with an expect size of $n_0=800$ in the 
trial and $n_1=9,200$ in the OS.  The RMSE for adaptive estimator is generally similar 
to that of the trial estimator.

Table 2 presents the pointwise coverage results.  When working models are correctly 
specified, the coverage of all estimators achieve approximately nominal coverage 
levels.  Under misspecified models, the OS estimator has extremely poor 
coverage that is sustained in large samples, whereas the trial estimator
maintains nearly ominal coverage.  The adaptive estimator achieves much improved 
coverage rates relative to the OS estimator and has close to nominal coverage
in large samples, as the estimator tends to shrink towards the trial estimator
to remove bias.

\begin{table}[htbp]
  \centering
  \scalebox{.605}{
    \begin{tabular}{lcc|cccccc|ccccccc}
      \multicolumn{3}{c}{} & \multicolumn{6}{c}{\textbf{Correctly Specified Models}} & \multicolumn{6}{c}{\textbf{Misspecified Models}} \\ \toprule
      &       &       & \multicolumn{2}{c}{\textbf{Trial}} & \multicolumn{2}{c}{\textbf{Obs Study}} & \multicolumn{2}{c}{\textbf{Adaptive}} & \multicolumn{2}{c}{\textbf{Trial}} & \multicolumn{2}{c}{\textbf{Obs Study}} & \multicolumn{2}{c}{\textbf{Adaptive}} \\
      & Trial Size & OS Size & Bias  & RMSE  & Bias  & RMSE  & Bias  & RMSE  & Bias  & RMSE  & Bias  & RMSE  & Bias  & RMSE \\
      \midrule
      \textbf{Pointwise} &       &       &       &       &       &       &       &       &       &       &       &       &       &  \\
      \quad 5th percentile  & 80/1,600 & 920/18,400 & -0.01 & 0.40  & 0.00  & 0.19  & 0.00  & 0.18  & -0.01 & 5.98  & -10.00 & 10.75 & -3.97 & 6.90 \\
      \quad 25th percentile & 80/1,600 & 920/18,400 & 0.00  & 0.33  & 0.01  & 0.12  & 0.01  & 0.13  & 0.04  & 4.18  & -4.65 & 4.98  & -2.16 & 3.90 \\
      \quad 50th percentile & 80/1,600 & 920/18,400 & 0.00  & 0.31  & 0.00  & 0.11  & 0.00  & 0.12  & 0.02  & 4.41  & -4.04 & 4.30  & -2.07 & 3.74 \\
      \quad 75th percentile & 80/1,600 & 920/18,400 & 0.00  & 0.36  & 0.00  & 0.11  & 0.01  & 0.13  & 0.11  & 6.08  & -9.05 & 28.44 & -2.97 & 5.12 \\
      \quad 95th percentile & 80/1,600 & 920/18,400 & 0.00  & 0.44  & 0.00  & 0.19  & 0.00  & 0.19  & 0.40  & 11.01 & -56.64 & 310.39 & -3.02 & 8.67 \\
      \textbf{Integrated} & 80/1,600 & 920/18,400 & 0.00  & 0.35  & 0.00  & 0.14  & 0.00  & 0.12  & 0.13  & 6.08  & -14.14 & 222.87 & -2.63 & 5.11 \\
      \midrule
      \textbf{Pointwise} &       &       &       &       &       &       &       &       &       &       &       &       &       &  \\
      \quad 5th percentile  & 400/1,600 & 4,600/18,400 & 0.00  & 0.19  & 0.01  & 0.10  & 0.01  & 0.10  & -0.04 & 2.48  & -11.48 & 11.67 & -0.99 & 3.53 \\
      \quad 25th percentile & 400/1,600 & 4,600/18,400 & 0.00  & 0.15  & 0.00  & 0.06  & 0.00  & 0.07  & 0.00  & 1.74  & -3.99 & 4.08  & -0.40 & 1.96 \\
      \quad 50th percentile & 400/1,600 & 4,600/18,400 & 0.00  & 0.15  & 0.00  & 0.06  & 0.00  & 0.07  & -0.01 & 2.12  & -3.99 & 4.07  & -0.45 & 2.31 \\
      \quad 75th percentile & 400/1,600 & 4,600/18,400 & 0.00  & 0.16  & 0.00  & 0.06  & 0.00  & 0.07  & -0.01 & 2.42  & -12.08 & 119.94 & -0.62 & 2.65 \\
      \quad 95th percentile & 400/1,600 & 4,600/18,400 & 0.00  & 0.21  & 0.00  & 0.10  & 0.00  & 0.10  & -0.26 & 4.80  & -622.84 & 9958.30 & -0.91 & 4.73 \\
      \textbf{Integrated} & 400/1,600 & 4,600/18,400 & -0.01 & 0.17  & 0.00  & 0.07  & 0.00  & 0.06  & 0.02  & 2.68  & -18.29 & 173.06 & -0.26 & 2.76 \\
      \midrule
      \textbf{Pointwise} &       &       &       &       &       &       &       &       &       &       &       &       &       &  \\
      \quad 5th percentile  & 800/1,600 & 9,200/18,400 & -0.01 & 0.14  & 0.00  & 0.07  & 0.00  & 0.07  & -0.05 & 1.82  & -12.03 & 12.14 & -0.23 & 2.16 \\
      \quad 25th percentile & 800/1,600 & 9,200/18,400 & 0.00  & 0.11  & 0.00  & 0.05  & 0.00  & 0.05  & -0.08 & 1.27  & -3.84 & 3.89  & -0.15 & 1.32 \\
      \quad 50th percentile & 800/1,600 & 9,200/18,400 & 0.00  & 0.12  & 0.00  & 0.04  & 0.00  & 0.05  & -0.05 & 1.61  & -4.04 & 4.09  & -0.13 & 1.65 \\
      \quad 75th percentile & 800/1,600 & 9,200/18,400 & 0.00  & 0.12  & 0.00  & 0.05  & 0.00  & 0.05  & -0.09 & 1.82  & -4.95 & 6.00  & -0.19 & 1.87 \\
      \quad 95th percentile & 800/1,600 & 9,200/18,400 & 0.00  & 0.15  & 0.00  & 0.07  & 0.00  & 0.07  & -0.25 & 3.47  & -139.49 & 984.86 & -0.40 & 3.50 \\
      \textbf{Integrated} & 800/1,600 & 9,200/18,400 & 0.00  & 0.12  & 0.00  & 0.05  & 0.00  & 0.05  & -0.11 & 2.00  & -29.26 & 586.80 & -0.17 & 2.05 \\
          \bottomrule
  \end{tabular}%
  }
  \caption{Pointwise bias and MSE of $\tauhat_1^r(\bv)$, $\tauhat_1^o(\bv)$, and 
  $\tauhat_1(\bv)$, evaluated at different $\bV$ (percentiles of $X_1$),
  in scenarios with correctly specified and misspecified working models.
  The integrated bias and MSE are the average performance over the support of 
  $\bV$.  The trial and OS sizes refer to expected sizes in training/validation
  based on 8\% of total sample being selected to trial, for total sizes of 
  $n=1000,5000,10,000$ and $n_{v}=20,000$.  The results are based on $R=1,000$ simulation replications.}
  \label{t:bias}%
\end{table}%

\begin{table}[htbp]
  \centering
  \scalebox{.805}{
    \begin{tabular}{lcc|ccc|ccc}
      \multicolumn{3}{c}{}       & \multicolumn{3}{c}{\textbf{Correctly Specified Models}} & \multicolumn{3}{c}{\textbf{Misspecified Models}} \\ \toprule
      & Trial size & OS size & Trial & OS    & Adaptive & Trial & OS    & Adaptive \\
      \midrule
      \textbf{Pointwise} &       &       &       &       &       &       &       &  \\
    \quad 5th percentile  & 80/1,600 & 920/18,400 & 92.1\% & 94.8\% & 95.9\% & 96.9\% & 28.1\% & 69.8\% \\
    \quad 25th percentile & 80/1,600 & 920/18,400 & 93.0\% & 94.3\% & 95.2\% & 95.9\% & 33.5\% & 71.4\% \\
    \quad 50th percentile & 80/1,600 & 920/18,400 & 93.6\% & 95.3\% & 95.6\% & 95.0\% & 22.6\% & 65.0\% \\
    \quad 75th percentile & 80/1,600 & 920/18,400 & 93.2\% & 95.1\% & 96.2\% & 94.9\% & 24.8\% & 65.9\% \\
    \quad 95th percentile & 80/1,600 & 920/18,400 & 94.2\% & 95.4\% & 95.4\% & 94.6\% & 92.7\% & 95.3\% \\
    \midrule
    \textbf{Pointwise} &       &       &       &       &       &       &       &  \\
    \quad 5th percentile  & 400/1,600 & 4,600/18,400 & 94.0\% & 96.0\% & 95.8\% & 98.2\% & 0.8\% & 89.3\% \\
    \quad 25th percentile & 400/1,600 & 4,600/18,400 & 94.6\% & 93.9\% & 95.0\% & 97.2\% & 0.8\% & 88.2\% \\
    \quad 50th percentile & 400/1,600 & 4,600/18,400 & 94.3\% & 95.1\% & 95.8\% & 95.8\% & 0.2\% & 86.3\% \\
    \quad 75th percentile & 400/1,600 & 4,600/18,400 & 95.0\% & 95.9\% & 97.1\% & 97.6\% & 4.1\% & 88.5\% \\
    \quad 95th percentile & 400/1,600 & 4,600/18,400 & 95.4\% & 94.0\% & 96.4\% & 97.9\% & 43.3\% & 97.0\% \\
    \midrule
    \textbf{Pointwise} &       &       &       &       &       &       &       &  \\
    \quad 5th percentile  & 800/1,600 & 9,200/18,400 & 95.0\% & 94.9\% & 96.2\% & 98.3\% & 0.1\% & 96.8\% \\
    \quad 25th percentile & 800/1,600 & 9,200/18,400 & 94.2\% & 94.4\% & 94.5\% & 97.6\% & 0.0\% & 96.0\% \\
    \quad 50th percentile & 800/1,600 & 9,200/18,400 & 93.6\% & 93.9\% & 95.8\% & 95.4\% & 0.0\% & 94.0\% \\
    \quad 75th percentile & 800/1,600 & 9,200/18,400 & 94.2\% & 95.5\% & 94.9\% & 96.3\% & 0.8\% & 95.0\% \\
    \quad 95th percentile & 800/1,600 & 9,200/18,400 & 95.2\% & 95.0\% & 96.3\% & 97.0\% & 32.8\% & 96.6\% \\
      \bottomrule
    \end{tabular}%
  }
    \caption{Empirical coverage rate of $\tau(\bv)$ by 95\% normal 
    approximation CIs based on estimating the asymptotic 
    variance of $\tauhat_z^r(\bv)$, $\tauhat_z^o(\bv)$, and $\tauhat_z(\bv)$, in 
    scenarios with correctly specified and misspecified working models. 
    The trial and OS sizes refer to expected sizes in training/validation
    based on 8\% of total sample being selected to trial, for total sizes of 
    $n=1000,5000,10,000$ and $n_{v}=20,000$. 
    The results are based on $R=1,000$ simulation replications.}
  \label{t:coverage}%
\end{table}%

\section{Analysis of Women's Health Initiative Data}
The WHI was a collection of studies that investigated strategies for preventing common sources of mortality and morbidity 
among postmenopausal women, including cancer, cardiovascular disease, and fractures \citep{anderson1998design}. Among its 
various components, the WHI included randomized trials and an OS examining the long-term health effects of hormone replacement therapy 
(HRT). In the following analysis, we focus on an analysis of the effects of estrogen plus progestin (E+P) combined hormone 
therapy on the 5-year risk of coronary heart disease (CHD) using data from the E+P trial 
and the cohort of women from the OS with a uterus who were either using either combined 
hormone therapy or no hormone therapy.  

The outcome $Y$ was a binary variable defined by locally and centrally adjudicated 
occurrences of myocardial infarction (MI) and death due to CHD within 5 years of follow-up.  
As the rates of early loss-to-follow-up were low ($<4.4$\% in our sample of trial and OS participants),
we estimated 5-year risks excluding participants without full 5-year follow-up under
the assumption of independent censoring.
The covariates $\bX$ included age, ethnicity (white or non-white), education status 
(not high school graduate, high school graduate but not college graduate, college graduate or higher), 
smoking status (ever vs. never), body mass index, a physical functioning score based 
on the SF-36 physical functioning subscore, previous occurrence of menopausal symptoms 
such as hot flashes or night sweats, and age at menopause. These covariates were 
selected to cover the basic adjustment variables used in many existing analyses 
of the WHI data \citep{prentice2005combined,prentice2006combined}. 
We limited patients in the E+P trial to those without any history of hormone therapy 
and patients in the OS cohort to those who either never had hormone therapy or are 
current users of E+P therapy with $<5$ years of current use, as the duration 
of hormone therapy use has been found to have a substantial impact on treatment 
effects \citep{prentice2005combined}. Among subjects in the E+P 
trial and the OS cohort, we further excluded subjects without outcome follow-up,
without data on HRT use at baseline, and with missing data in any $\bX$. The final 
analytic dataset included $9,825$ subjects in the trial and $29,260$ in the OS. 

An initial analysis based on fitting Cox proportional hazards models that adjusted
for $\bX$ as main effects and included subjects with $<5$ years of follow-up, 
separately in the trial and OS subjects, indicated a disrepancy in the estimated
treatment effects of E+P.  The adjusted hazard ratio for E+P vs. no HRT was estimated to be 
$1.225$ ($p=.081$) in the trial and $0.750$ ($p=.052$) in the OS.  However, the
adjusted hazard ratio is an overall effect that may obscure treatment effects
at specific patient profiles $\bX=\bx$.  We subsequently implemented the proposed
method to estimate CATEs  
given $\bV$ in the OS population, $\tau_1(\bv)$, where $\bV$ are the percentiles
of the estimated treatment differences between HRT vs. non-HRT given $\bX$.  We obtained
treatment differences by fitting separate logistic regressions of $Y\sim\bX$
in subjects receiving E+P vs. having no HRT the pooled trial and OS data and taking the differences
of predicted probabilities under each model.  The percentiles were obtained by
applying the empirical cumulative distribution to these estimated differences.
We $\tau_1(\bv)$ over an evenly spaced
grid in the interval $(0,1)$ for $\bv$. The total data was split such that a random 
20\% was used training and 80\% for validation.  The final estimates are estimated with
both the training and validation sets after selecting $\lambda_n$.

\subsection{Results}

During the validation, the estimated MISE increased with $\lambda_n$ (Figure 1a),
that the trial estimator had worse performance than a combined estimator across
the support of $\bV$ in the validation data $\Dscr_v$.
Given that the optimal $\lambda_n$ was minimal, the resulting weights $\etahat$
coincide with estimates of the inerse-variance weights $\etastr$ that optimize
asymptotic variance irrespective of the potential bias in OS estimates.  The specific
estimates of $\etahat$ varied across $\bv$ (Figure 1b), with $\etahat>.9$ for 
$\bv\in (0.1,0.5)$, and with attenuated weights for $\bv >.5$.  These results suggest
that subjects with low estimated treatment differences in the trial and OS exhibited
similar treatment differences, whereas subjects with high treatment differences
exhibited various degrees of discrepancy.  As a result, the weights are shrunk
towards zero so that the adaptive estimates are shrunk towards the trial estimates.

The estimates of the CATE show there is heterogeneity of treatment effects across
the predicted treatment difference score $\bV=\bv$, although there is ambiguity due to differences in trial and
OS point estimates.  The trial estimates indicate there are minimal treatment differences
for participants with low scores and possibly elevated risk of E+P for participants
with high scores, though the estimates have high uncertainty due to the limited sample 
size.  On the other hand, the OS estimates indicate there are some significant
protective effects of E+P, based on the 95\% CIs, for participants with low scores 
and uncertain or possibly harmful effects for those with high scores.  The adaptive 
estimator combines these estimates while guarding against bias introduced from 
leveraging OS data, and indicates that there is a mild protective effect of E+P for 
those with low scores and uncertain or harmful effects for those with high scores.

\begin{figure}[h!]
  \centering
  \includegraphics[scale=.2]{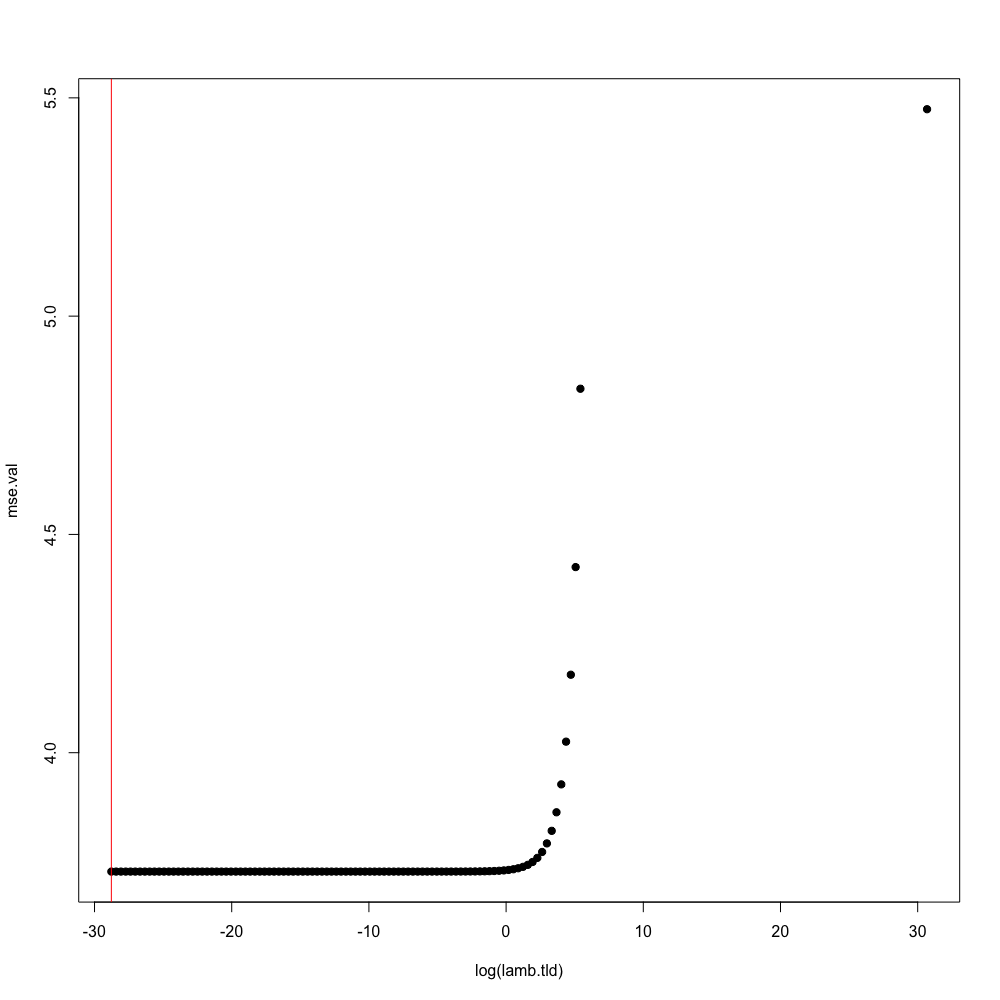}
  \includegraphics[scale=.2]{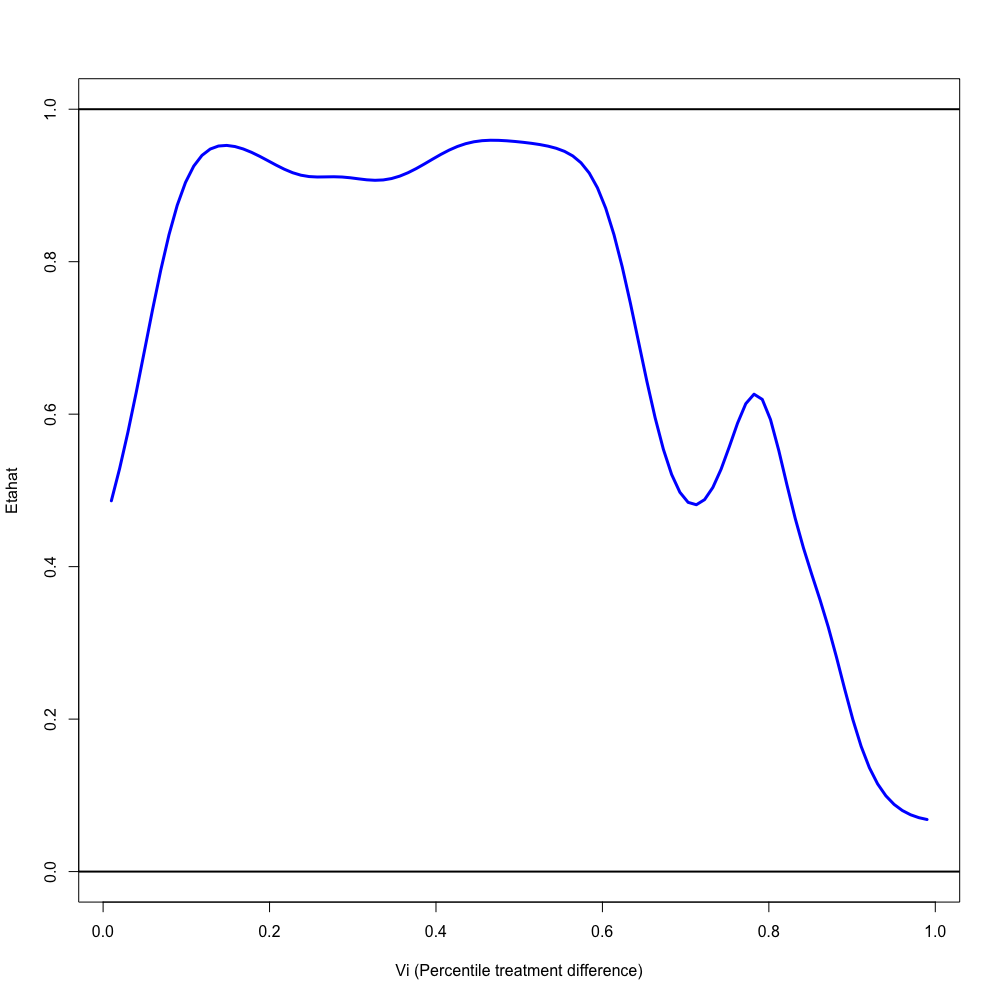}
\caption{
    Scaled MISE estimates $R_1(\lambda)$ in the validation data $\Dscr_v$ over grid of $\lambda_n$ (left). Estimated 
    adaptive weights $\etahat$ given $\bV$ (right).}
\end{figure}

\begin{figure}[h!]
  \centering
  \includegraphics[scale=.35]{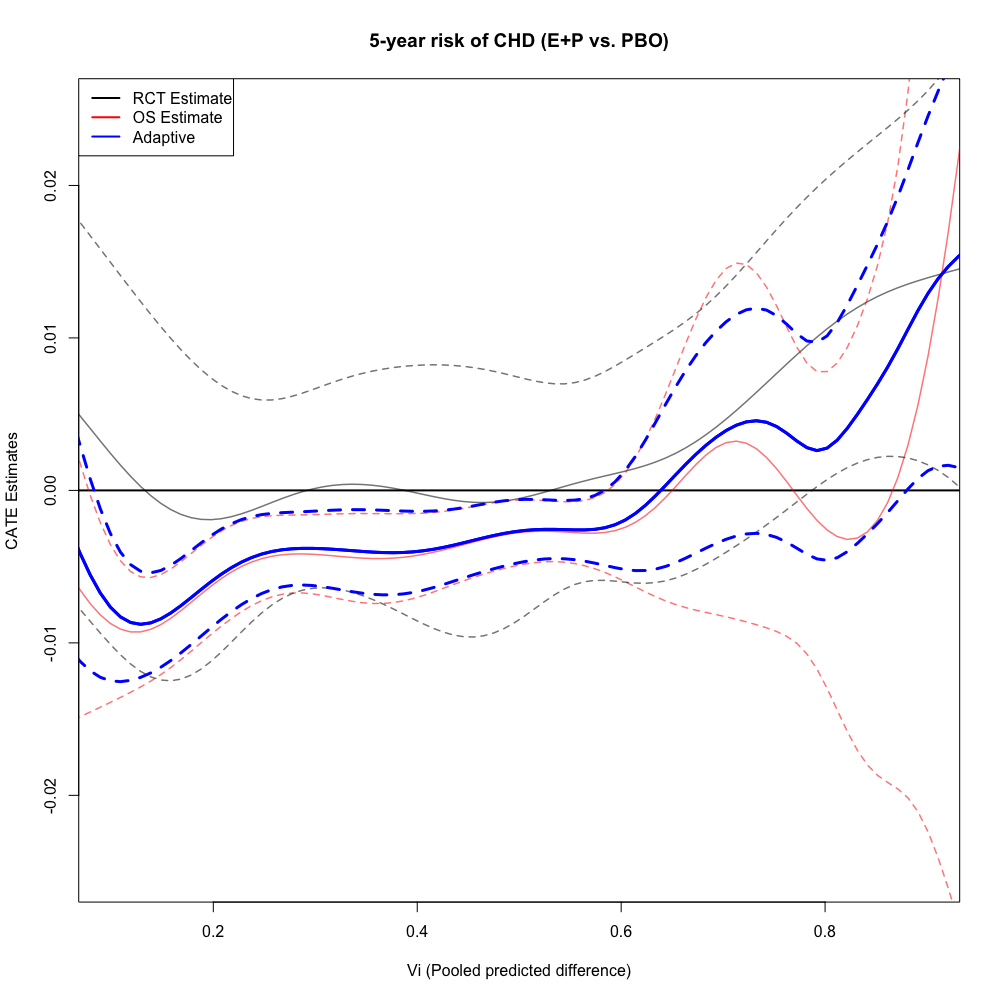}
\caption{
    Estimates of the 5-year risk of CHD using $\tauhat_z^r(\bv)$, $\tauhat_z^o(\bv)$, 
    and $\tauhat_z(\bv)$. Pointwise 95\% CI's are constructed based on estimates of 
    the asymptotic variance estimated from the influence functions.}
\end{figure}

\section{Discussion}

In this paper, we proposed an adaptive approach to combine CATE estimates from
parallel trial and OS data to mitigate bias due to confounding and treatment effect
heterogeneity and optimize efficiency when incorporating OS data.  The combination 
smoothly weights toward the trial estimate when bias remains after adjustment when 
discrepancies are found between the two estimates or otherwise combines estimators 
for efficiency.  This allows the data to directly assess the bias of the OS estimator 
and guide the decision of whether to combine estimates.  In large samples, the 
estimator is constructed to have a high probability of making the right decision
with respect to whether or not to combine estimates and also estimate optimal weights
for efficiency.

The proposed approach achieves adaptivity by anchoring the combined estimator on 
the trial estimator, under the assumption that the trial estimator always allows 
for consistent estimation of a relevant conditional causal effect. In practice, 
poor design or execution of trials may undermine the validity of RT estimates and
well-conducted OS can sometimes even produce results that are more trustworthy than 
poorly-conducted trials.
In these and other situations, it would be important to consider whether 
observed discrepancies between RT and OS estimates warrants confidence in the RT estimates. 
Relatedly, in practice no OS estimator can be expected to be exactly 
free of bias, and it may be unclear when the estimators are allowed to combine for 
efficiency. But just as misspecified parametric models provide useful approximations, 
assumptions such as no confounding condition in Assumption 2 serve as a useful guide 
for identifying cases in which combination should be allowed. In finite samples, 
we expect $\lambdahat_n$ to be $0$ or small when the bias fall below certain 
limits relative to their variability, though it would be interesting 
to theoretically consider the impact of the true order of the bias on the results.

We considered estimating the CATE over a low-dimensional reduction of $\bX$, $\bV=g(\bX)$ 
with the understanding that there is often interest in how treatment effects vary
over specific variables observed among $\bX$, while the role of other covariates
would to be adjust for confounding.  If there is interest in estimating the CATE
over many covariates, the proposed approach could be modified to adopt a parametric
specification for the CATE rather than estimating it nonparametrically.  Alternatively,
one could smooth over a score variable, in which an initial first-stage estimator
for the CATE is used for $\bV$, as done in the WHI example.  In general,
there may be multiple randomized and observational studies available for analysis.
One direction of future work is to extend the framework estabalished here to this case.  
Additionally, the size of the trial and OS data can be drastically different, and a more refined
theoretic analysis would allow for regimes in which $n_0$ and $n_1$ diverge at different 
rates.

\newpage

\appendix

\bibliography{sample}

\end{document}